\documentclass[a4paper,11pt]{article}
\usepackage{pos}

\usepackage{float}
\usepackage{color}
\usepackage{hyperref}
\usepackage{physics}
\usepackage[absolute,overlay]{textpos}

\title{LapH interpolating fields with open boundary conditions}
\ShortTitle{LapH with open boundary conditions}

\author[a]{Michele Della Morte}
\author*[a]{Olmo Francesconi}
\author[b]{Justus Tobias Tsang}

\affiliation[a]{CP3-Origins and IMADA, University of Southern Denmark, Campusvej 55, 5230 Odense M, Denmark}
\affiliation[b]{Theoretical Physics Department, CERN, 1211 Geneva 23, Switzerland}

\emailAdd{francesconi@imada.sdu.dk}

\abstract{The stochastic Laplacian Heaviside (LapH) method has proven to be
  successful in hadronic calculations. In this work, with charm--light
  spectroscopy in mind, we set up and optimise the LapH procedure limiting
  ourselves to the evaluation of two--point mesonic correlation functions. The
  calculations are performed on CLS ensembles with $N_f=2+1$ Wilson-Clover
  fermions on a $32^3\times64$ lattice with open boundary conditions. We analyse
  the interplay between the LapH parameters and the boundary effects, and
  implement a fitting procedure to isolate excitations coming from the border.

  \begin{textblock}{20}(15.0,1.70)
    CERN-TH-2022-212\\
  \end{textblock}% 
}

\FullConference{%
The 39th International Symposium on Lattice Field Theory,\\
8th-13th August, 2022,\\
Rheinische Friedrich-Wilhelms-Universit{\"a}t Bonn, Bonn, Germany
}

\begin{document}
\maketitle

%%%%%%%%%%%%%%%%%%%%%%%%%%%%%%%%%%%%%%%%%%%%%%%%%%%%%%%%
%%%%    introduction                                %%%%
%%%%%%%%%%%%%%%%%%%%%%%%%%%%%%%%%%%%%%%%%%%%%%%%%%%%%%%%
\section{Introduction}
In this work, we analyse the behaviour of the Laplacian-Heaviside (LapH)
\cite{HadronSpectrum:2009krc} smearing procedure in the presence of open boundary conditions. By
evaluating the pseudoscalar meson two--point functions we determine the portion of
the lattice unaffected by the boundary and the optimal LapH setup.

Our aim is the study of semileptonic decays in this framework, a largely
unexplored setup \cite{Boyle:2019cdl}. The evaluation of three--point
functions with local current insertions gives access to various quantities
relevant to flavour physics. The combination of theory predictions of
semileptonic form factors with experimental data gives access to components
of the Cabibbo-Kobayashi-Maskawa (CKM) matrix \cite{Cabibbo:1963yz, Kobayashi:1973fv}, where an
increase in precision could lead to hints of new physics. Also, the present
tension in lepton flavour universality \cite{London:2021lfn} makes the study of
semileptonic processes particularly interesting.

One source of uncertainty which limits the attainable precision of current
theory predictions arises from the trade-off between the contamination due to
excited states in the two-- and three--point functions and the increased
statistical uncertainty when their separation increases. Short separations
between source and sink make the evaluation of plateau in the relevant
quantities a delicate task and typically require sophisticated fitting
procedures. The LapH smearing procedure ameliorates the effects of excited
states on the relevant correlation functions.

Ensembles with open boundary conditions pave the path towards finer lattice
spacings, which are particularly relevant for heavy--light flavour
phenomenology. However, this requires to control any effects that might arise
from the open boundaries, which we aim to explore in this work.

%%%%%%%%%%%%%%%%%%%%%%%%%%%%%%%%%%%%%%%%%%%%%%%%%%%%%%%%
%%%%    LapH                                        %%%%
%%%%%%%%%%%%%%%%%%%%%%%%%%%%%%%%%%%%%%%%%%%%%%%%%%%%%%%%
\section{Computation of meson two--point functions}
Our computations are based on the recently developed LapH \cite{HadronSpectrum:2009krc} smearing
procedure implemented on each time--slice by the smearing matrix
\begin{equation}
 \mathcal{S}(\bar{x},\bar{y}) = \Theta \left( \sigma_s^2 + \Delta \right) \simeq V_s V_s^\dagger,
\end{equation}
where $V_s$ is the matrix containing columns of the eigenvectors associated with
the $N_{ev}$ lowest-lying eigenvalues of the 3D Laplacian $\Delta$. The
smeared quark fields thus take the form
\begin{equation}
 \tilde{\chi} = \chi\mathcal{S} = \bar{\psi}\gamma_4 \mathcal{S}, \qquad \tilde{\psi} = \mathcal{S} \psi.
\end{equation}
Then, by introducing some $Z(N)$ noise vectors $\rho$ in the LapH eigenspace and
by defining the related dilution projector $\mathcal{P}$ in the time, spin and
$ev$ space \cite{Morningstar:2011ka} one can identify two new objects: quark sinks
$\varphi = \mathcal{S}\Omega^{-1}V_s\mathcal{P}\rho$ and quark sources
$\varrho=V_s\mathcal{P}\rho$, where only the former requires a Dirac matrix
inversion to be computed.  With these building blocks, the meson correlation
functions take the form
\begin{align}
    C(t - t_0)  &= \langle 
                  \tilde{\chi}_{a}(t) \Gamma^A \tilde{\psi}_{a}(t) \
                  \tilde{\chi}_{b}(t_0) \Gamma^B \tilde{\psi}_{b}(t_0)
                  \rangle
                  = \langle \Gamma^B \ Q_{ba}(t_0, t) \ \Gamma^A \ Q_{ab}(t,t_0) \rangle\\
                & = \langle \Gamma^B \varphi_{b}(t_0)\varrho_{a}(t)^* \Gamma^A \varphi_{a}(t)\varrho_{b}(t_0)^* \rangle
                  = \langle \Gamma^B \bar{\varrho}_{b}(t_0)\bar{\varphi}_{a}(t)^* \Gamma^A \varphi_{a}(t)\varrho_{b}(t_0)^* \rangle\\
                & = \langle \mathcal{M}^{\Gamma^A}(\bar{\varphi}, \varphi, t) \ \mathcal{M}^{\Gamma^B}(\bar{\varrho}, \varrho, t_0)^* \rangle.
\end{align}
Where $\mathcal{M}$ are the (LapH subspace-sized) diluted meson fields. In this
work we are interested only in evaluating pseudoscalar mesons, therefore we will
study only the $A_{1u}^{(+)}$ channel of the octahedral symmetry
group. Moreover, we are going to employ a trivial $Z(1)$ noise vector, and no
dilution scheme, therefore restricting ourselves to \textit{exact distillation}.

Due to the open boundary conditions, and up to leading order in chiral
perturbation theory, we expect the two--point functions to fall off as
\cite{Luscher:2012av}
\begin{equation}\label{eq:sinh}
 C(t) \propto \sinh (m (\tilde{T} - t)),
\end{equation}
where $\tilde{T}$ is a free parameter. In particular, it defines a “virtual”
lattice border inside the lattice itself and can be easily obtained via a fit.

\section{Results}

%%%%%%%%%%%%%%%%%%%%%%%%%%%%%%%%%%%%%%%%%%%%%%%%%%%%%%%%
%%%%    Simulation setup                           %%%%
%%%%%%%%%%%%%%%%%%%%%%%%%%%%%%%%%%%%%%%%%%%%%%%%%%%%%%%%

Measurements are performed on configurations generated by the CLS initiative
\cite{Bruno:2014jqa} with $N_f=2+1$ non-perturbatively improved Wilson fermions with open
boundary conditions, details of which are given in
Table~\ref{tab:ensembleparameters}.
\begin{table}
    \begin{tabular}{ccccccccc}
    $\mathrm{id}$   & $a[\mathrm{fm}]$  & $N_s$ & $N_t$ & $\kappa_l$      & $\kappa_s$        & $\kappa_c$ \cite{Gerardin:2019rua}  & $m_\pi [\mathrm{MeV}]$ & $m_K [\mathrm{MeV}]$\\ \hline
    $B105$ & $0.086$ & $32$  & $64$  & $0.136970$ & $0.13634079$ & $0.123244(19)$                & $280$         & $480$
    \end{tabular}
    \caption{Relevant parameters for the lattice ensemble studied in this work.}
    \label{tab:ensembleparameters}
\end{table}
Gauge averages are extracted from a subsample of $100$ configurations and error
estimates are obtained from the bootstrap procedure. On each configuration, we
generate the LapH subspace with $96$ eigenvectors from which we compute the
pseudoscalar meson fields at three source positions $t_0 = 9, 15, 31$.
Since we are using exact distillation, the meson correlation functions with
fewer numbers of eigenvectors can be obtained without any additional
computations by considering the $4 N_{ev} \times 4 N_{ev}$ square sub-matrix
for each of the meson field matrices.

%%%%%%%%%%%%%%%%%%%%%%%%%%%%%%%%%%%%%%%%%%%%%%%%%%%%%%%%
%%%%    eigenvalues                                 %%%%
%%%%%%%%%%%%%%%%%%%%%%%%%%%%%%%%%%%%%%%%%%%%%%%%%%%%%%%%
\subsection{Boundary effect on the LapH eigenspace}
\begin{figure}[t]
 \centering
 \makebox[\textwidth][c]{\includegraphics[width=1.2\textwidth]{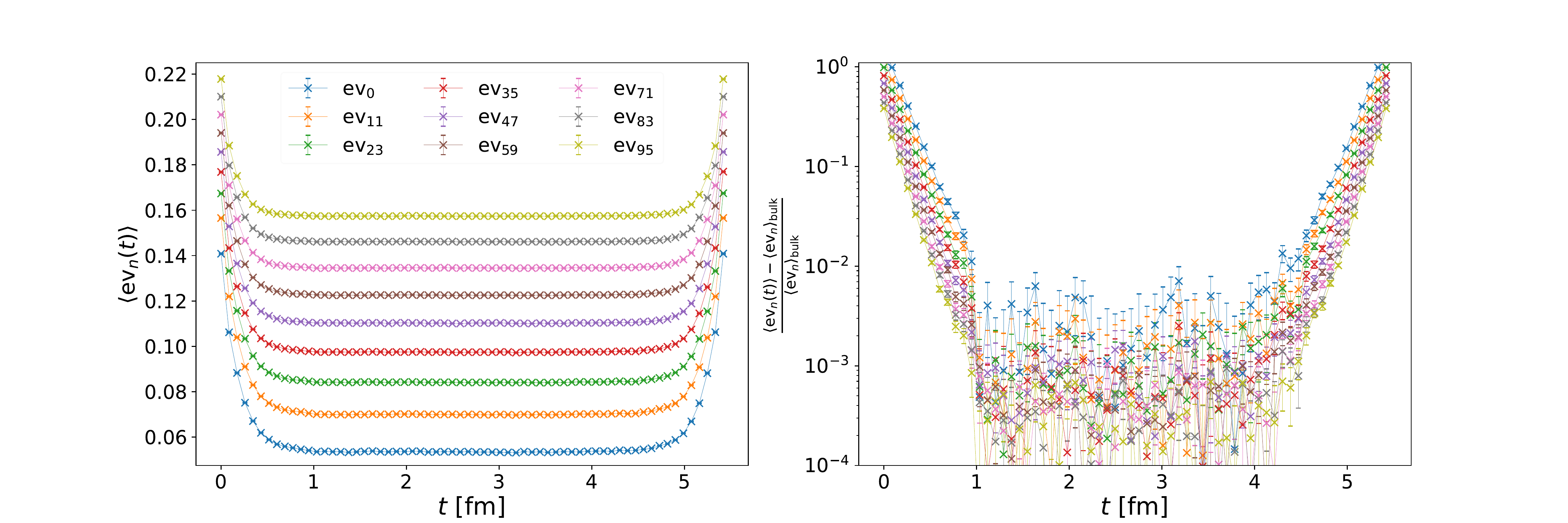}}
 \caption{\textbf{Left}: eigenvalues of the Laplacian as a function of the
   lattice time coordinate. \textbf{Right}: deviation from the bulk value for
   each eigenvalue.}
 \label{fig:ev}
\end{figure}

As a first step to estimate the effect of the open boundary conditions on the
LapH subspace we consider the gauge average the LapH eigenvalues as a function
of the lattice time coordinate. Periodic boundary conditions guarantee
translational invariance of the LapH subspace, while with open boundary
conditions this might not be the case. As can be seen from the left--hand panel
of figure~\ref{fig:ev}, we observe that the eigenvalues are time--independent
provided they are sufficiently far from the boundary. For the ensemble under
consideration this amounts to approximately 12 time--slices ($\sim
1\,\mathrm{fm}$). Taking the bulk value to be the average of the central 24
time--slices, the right--hand panel of the same figure shows the relative
deviation from this for each eigenvalue. We find that the deviation from the
bulk value is to a good approximation exponential.

%%%%%%%%%%%%%%%%%%%%%%%%%%%%%%%%%%%%%%%%%%%%%%%%%%%%%%%%
%%%%    imaginary correlator                        %%%%
%%%%%%%%%%%%%%%%%%%%%%%%%%%%%%%%%%%%%%%%%%%%%%%%%%%%%%%%
\subsection{Considerations on the realness of the meson correlation function}
Another important aspect of the computation of the two--point functions with
open boundary conditions concerns the effects that boundary vacuum states have
on the correlation function. Following~\cite{Guagnelli:1999zf} where the
analogous discussion for the Schr\"odinger functional case is presented, we
write a two--point function with open boundary conditions in the quantum
mechanical representation as
\begin{equation}
    C_2(t_1,t_2)= \frac{1}{Z}  \langle i_0| P(t_1) P^\dagger(t_2) |i_0\rangle \;, \quad t_1>t_2\;, 
\end{equation}
where $| i_0 \rangle$ is the state at the boundaries, with vacuum quantum
numbers. In general
\begin{equation}
    | i_0 \rangle = w_0 |0\rangle + w_1|0'\rangle + \cdots
\end{equation}
where $|0\rangle$ is the lowest eigenstate of the Hamiltonian $H$ and
$|0'\rangle$ is the first excited vacuum state. Note that in this formulation
the overlap factors $w_i$ are in general complex and the partition function
reads $\langle i_0 | e^{-TH} | i_0 \rangle$, where $T$ is the temporal extent of
the lattice. Now, up to a constant, the two--point function in the Heisenberg
picture is
\begin{equation}
    C_2(t_1,t_2) \propto \langle i_0 |e^{-(T-t_1)H} P e^{-(t_1-t_2)H}P^\dagger e^{-t_2H} |i_0 \rangle\;.
\end{equation}
We consider the different contributions starting with the vacuum-vacuum
contribution to $C_2(t_1,t_2)$
\begin{equation}
    w_0^* w_0 \langle 0| P e^{-(t_1-t_2)H}P^\dagger|0\rangle \;,
\end{equation}
which, under the assumptions above, is real. In order to simplify the discussion
we assume that $P$ excites just one state out of $|0\rangle$ and a different one
out of $|0'\rangle$. The two ground state vacuum to first excited state vacuum contributions read
\begin{equation}
    w_0 w_1^* \langle 0'|e^{-(T-t_1)E_{0'}} P e^{-(t_1-t_2)H}P^\dagger|0\rangle \;, \textrm{ and} \quad
    w_0^*w_1 \langle 0| P e^{-(t_1-t_2)H}P^\dagger e^{-E_{0'}t_2}|0'\rangle \;.
\end{equation}
Even within the strong assumptions above their sum is not real.

However, as long as only pure QCD is taken into account, a single meson state
can be represented equivalently with the two charge conjugated flavour
combinations, and the resulting correlation functions are complex conjugates
of each other. If the boundary contaminations are small, the real part of the
correlation functions still correctly represents the meson state.

%%%%%%%%%%%%%%%%%%%%%%%%%%%%%%%%%%%%%%%%%%%%%%%%%%%%%%%%
%%%%    correlators                                 %%%%
%%%%%%%%%%%%%%%%%%%%%%%%%%%%%%%%%%%%%%%%%%%%%%%%%%%%%%%%
\subsection{Boundary effects on the meson correlation functions}
Now we consider the results for the two--point functions computed with the
machinery described above. In the following we restrict ourselves to the
evaluation of the correlation functions for $\pi$, $K$ and $D_s$ pseudoscalar
mesons, and the corresponding effective masses computed as 
$m(t + \flatfrac{1}{2}) = \log(\flatfrac{C(t)}{C(t+1)})$.
To ensure that we have full control over all contributions to the correlation
function, we want to understand and disentangle the effects arising from the
LapH parameters and those coming from the boundaries.

\begin{figure}
    \centering    
    \includegraphics[width=0.32\textwidth]{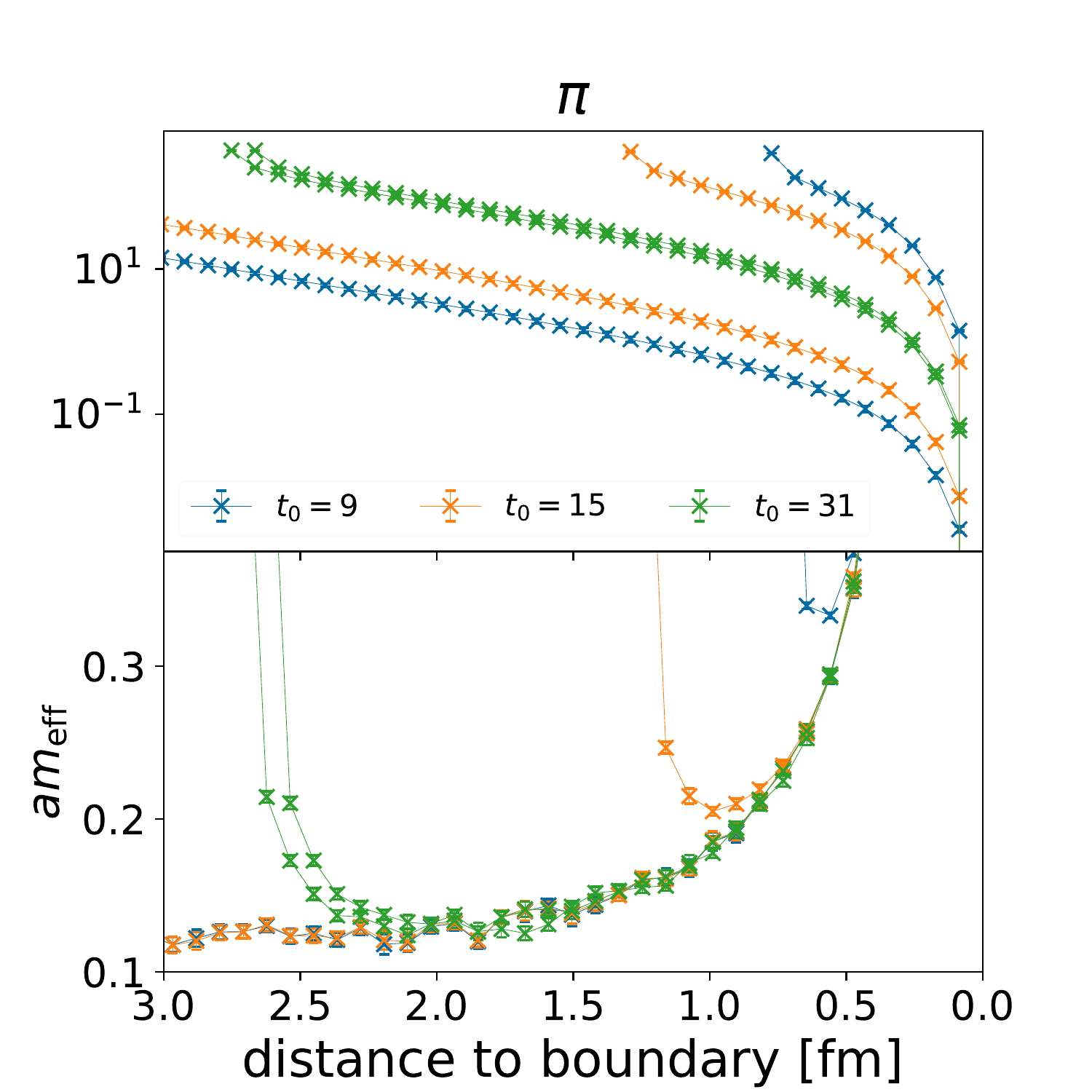} 
    \includegraphics[width=0.32\textwidth]{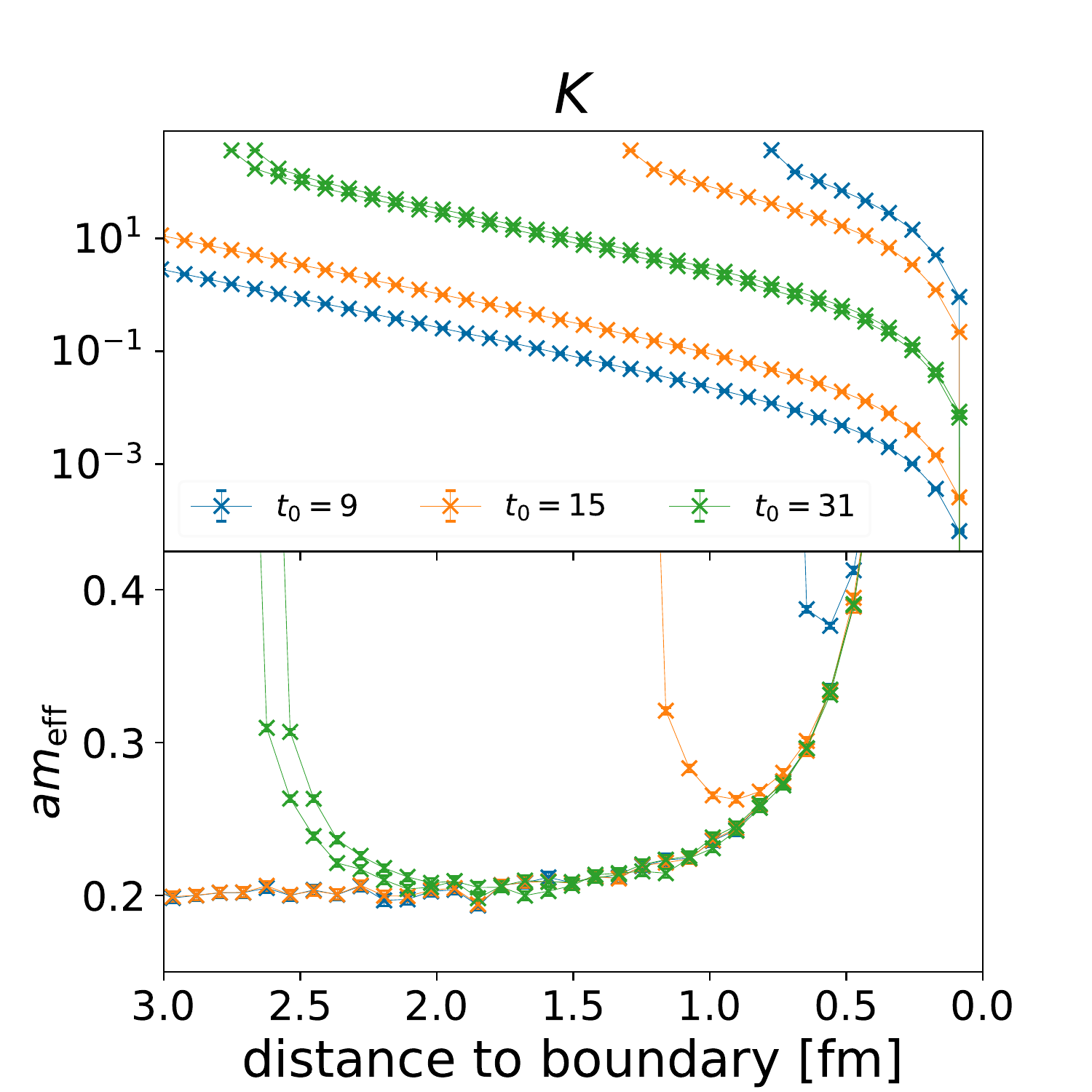}
    \includegraphics[width=0.32\textwidth]{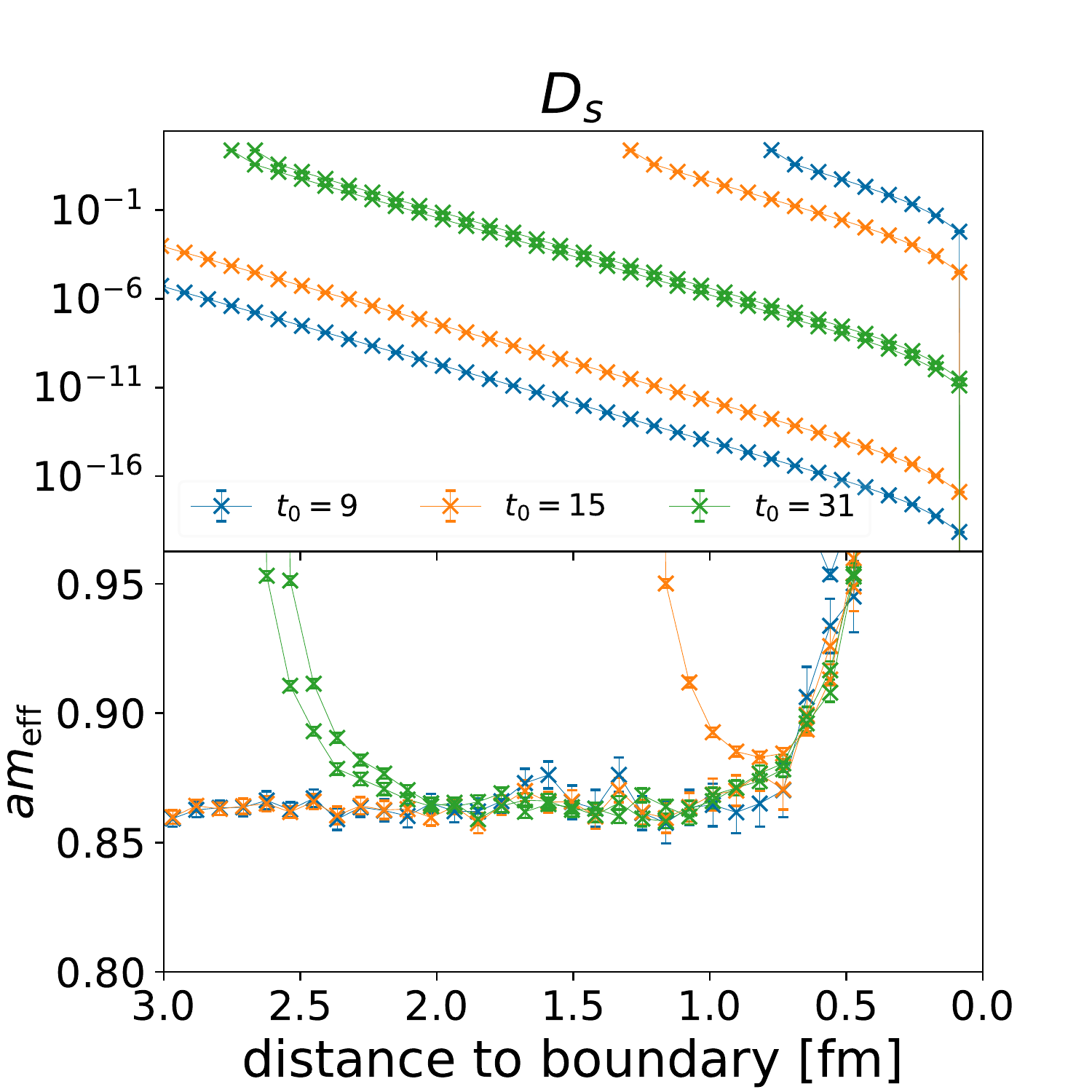}
    \caption{Comparison of meson correlation functions for the pion (left), kaon
      (middle) and the $D_s$ (right) computed at source times $t_0 = 9, 15, 31$
      and with $N_{ev} = 96$. The top panels show the correlation functions, the
      bottom panels the effective mass.}
    \label{fig:corr_1}
\end{figure}

In figure~\ref{fig:corr_1} we plot the two--point functions as a function of the
distance from the nearest boundary. The boundary effects are clearly visible at
the right hand edges of the various panels. The fact that the different branches
all coincide close to the boundary shows that these effects are independent of
the source position. They correspond to a tower of excitations arising from the
boundary. In the cases where sufficiently many time slices are available and at
this level of statistical uncertainty, we observe the onset of a plateau at
$\sim 2.0, 1.8, 1.0$ fm from the boundary for $\pi, K$ and $D_s$, respectively.

\begin{figure}
    \centering
    \includegraphics[width=0.32\textwidth]{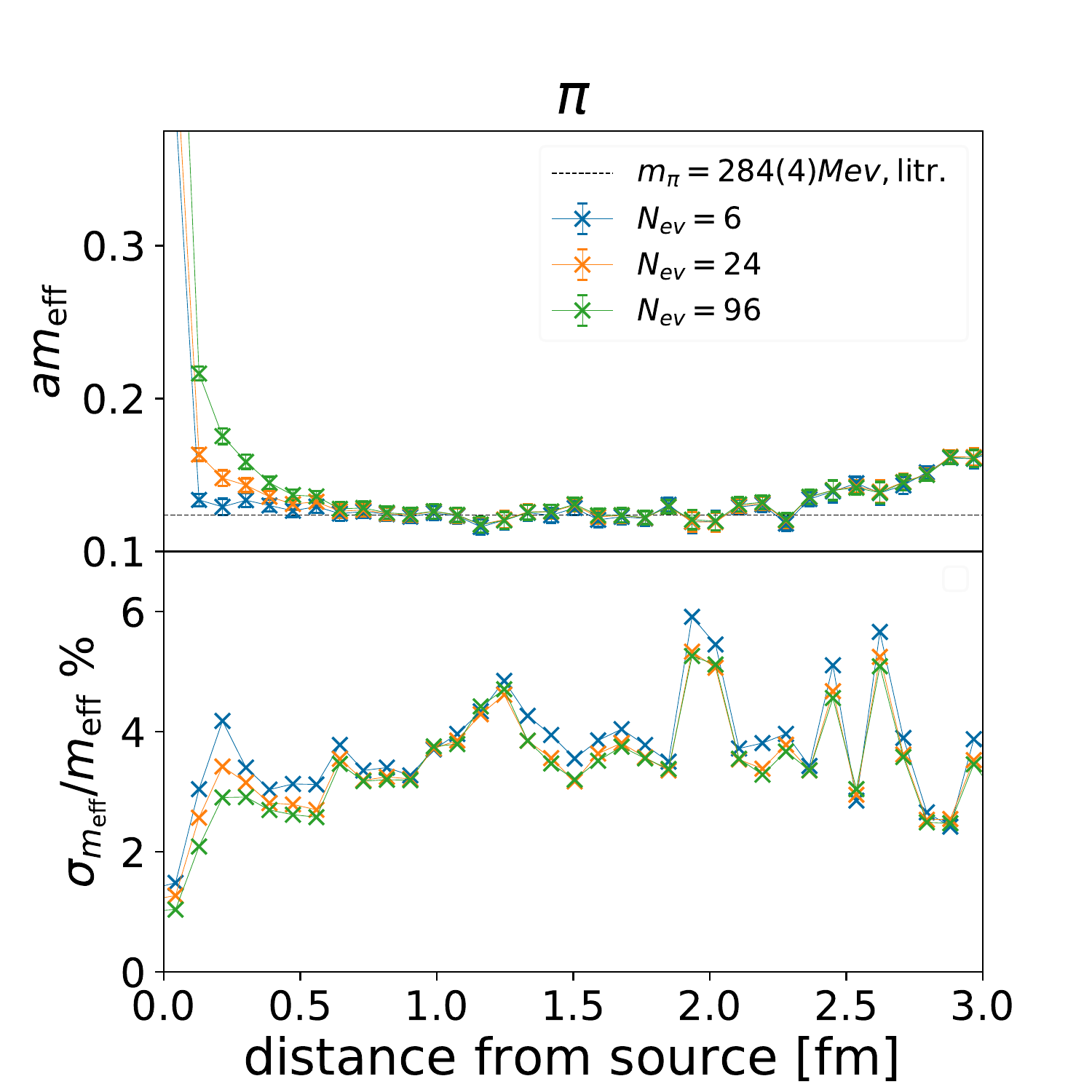} 
    \includegraphics[width=0.32\textwidth]{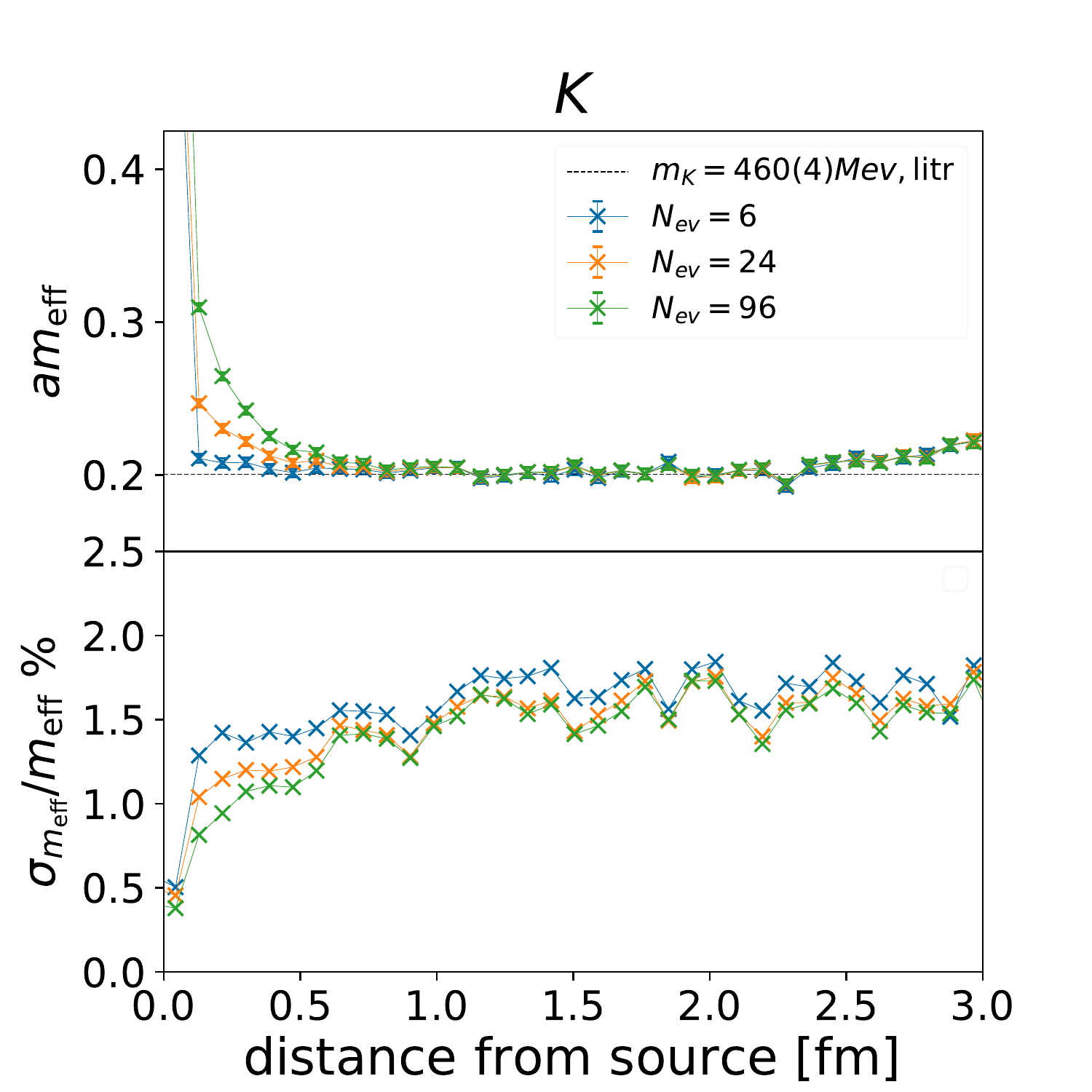}
    \includegraphics[width=0.32\textwidth]{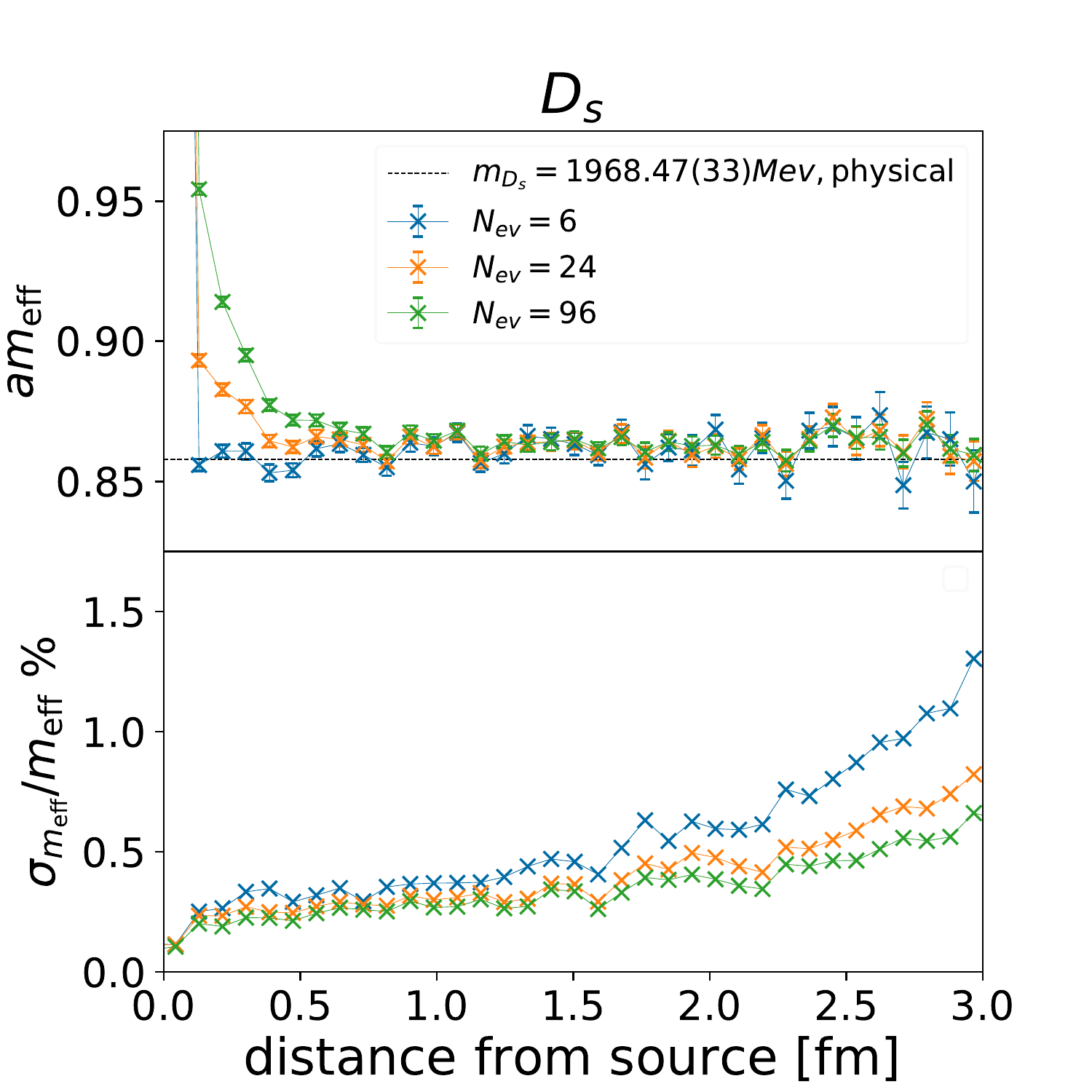}
    \caption{Comparison of the meson effective masses (top) and the relative
      uncertainties (bottom) for varying number of eigenvectors $N_{ev} = 6, 24,
      96$. The correlation functions are computed for one fixed source at $t_0 =
      15$ and we restrict ourselves to the forward branch.}
    \label{fig:corr_2}
\end{figure}

In figure~\ref{fig:corr_2} we restrict ourselves to a single source position
$t_0=15$ and investigate how quickly the effective mass approaches the plateau
as a function of the number of eigenvectors used in the construction of the LapH
subspace. As expected, a smaller number of lowest-lying eigenvectors results in
a broader smearing and less contamination by excited states near the source.
We observe a moderate increase in the statistical uncertainty as the number of
eigenvectors is reduced. In our setup we estimate the optimal number of
eigenvalues to be around $24$.

\begin{figure}
    \centering
    \includegraphics[width=0.32\textwidth]{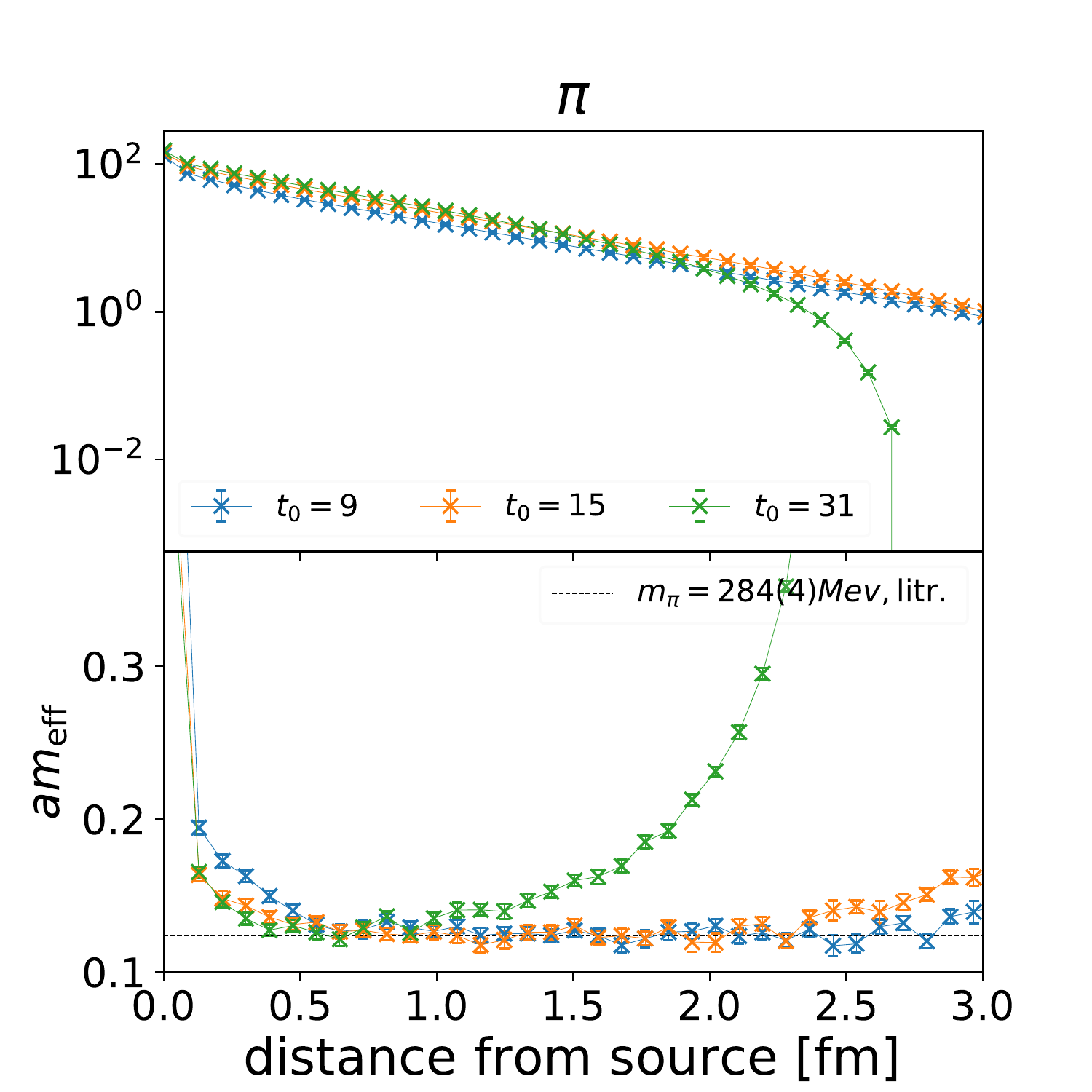}
    \includegraphics[width=0.32\textwidth]{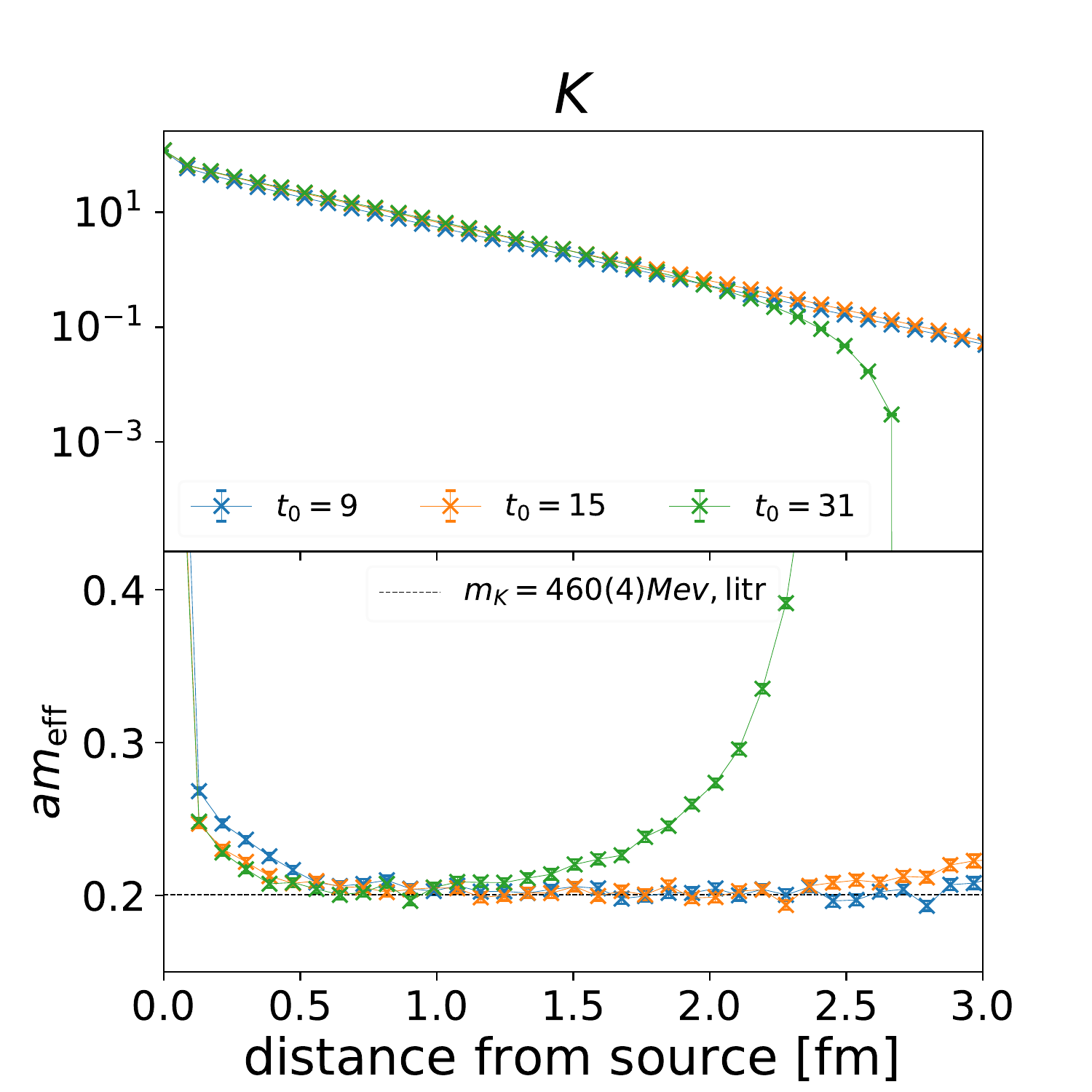}
    \includegraphics[width=0.32\textwidth]{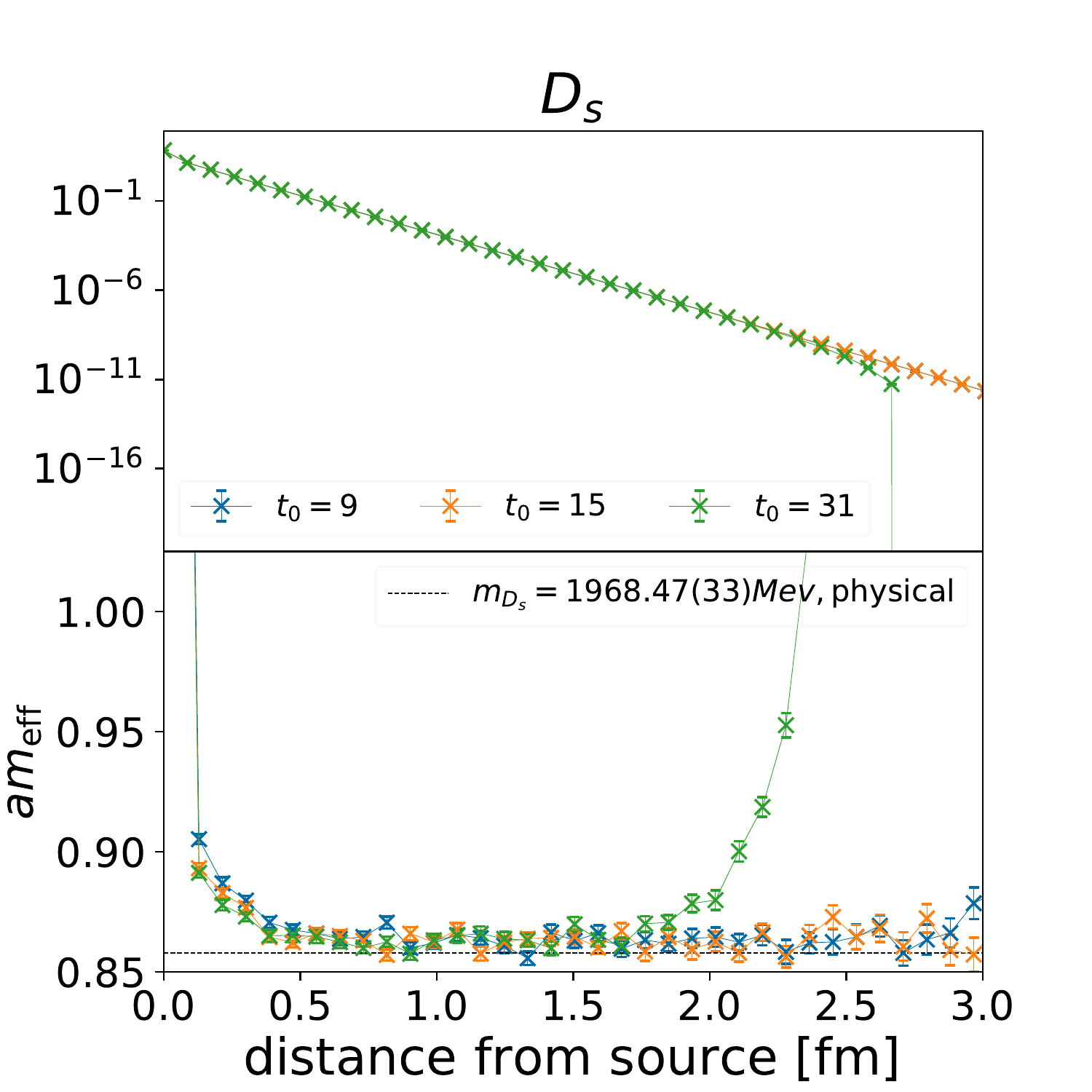}
    \caption{Comparison of the forward branch of the meson correlation functions
      (top) and effective masses (bottom) computed at different source times
      $t_0 = 9, 15, 31$ and with $N_{ev} = 24$.}
    \label{fig:corr_3}
\end{figure}

Lastly, in figure~\ref{fig:corr_3} we show the forward--time correlation
functions for $N_{ev} = 24$ and compare the short time--separation behaviour for
three source positions. The asymptotic value to which the effective mass
approaches is compatible with the literature~\cite{Bruno:2014jqa} (for the case of $\pi$
and $K$) or in good agreement with the physical value (for the $D_s$). However,
for the case of the pion and kaon at small time separations from the source, we
observe that the $t_0=9$ data set deviates from the other two source
positions. Crucially, this contamination does not arise from the farther
boundary, but rather from the one close to the source. We conclude that this
amounts to a mixing between the excited states generated at the source and those
generated at the boundary.

However, as shown in figure~\ref{fig:corr_1}, the boundary effects are a global
characteristic of the lattice to which we have access by considering the large
time separation behaviour of the meson correlation functions. In particular, the
correlation functions are well described by eq.~\ref{eq:sinh} and the parameters
$m$ and $\tilde{T}$ can be determined from a fit. Subsequently, we can define a
continuous estimate of the effective mass of the meson state as
\begin{equation}
    m_{\sinh}(t) \equiv \dv{t} \log \left( \sinh (m (\tilde{T} - t)) \right) = m \ \coth(m(\tilde{T} - t)).
\end{equation}
We define the correction term $m_\mathrm{corr}$ by subtracting the fitted meson
mass $m$ from the quantity $m_\mathrm{sinh}(t)$, i.e. 
\begin{equation}
    m_{\textrm{corr}}(t) = m_{\sinh}(t) - m.
\end{equation}

\begin{figure}
    \centering
    \includegraphics[width=0.32\textwidth]{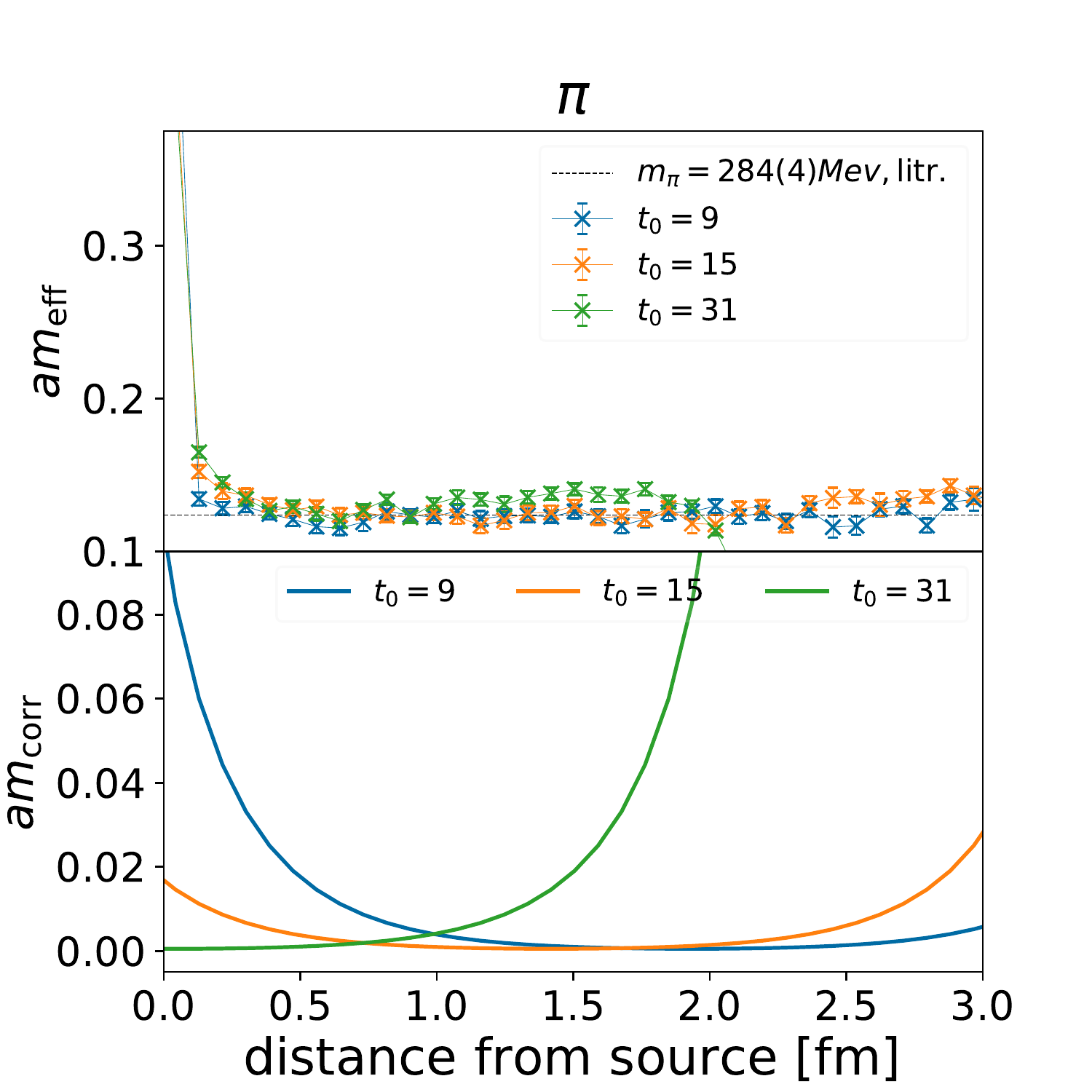}
    \includegraphics[width=0.32\textwidth]{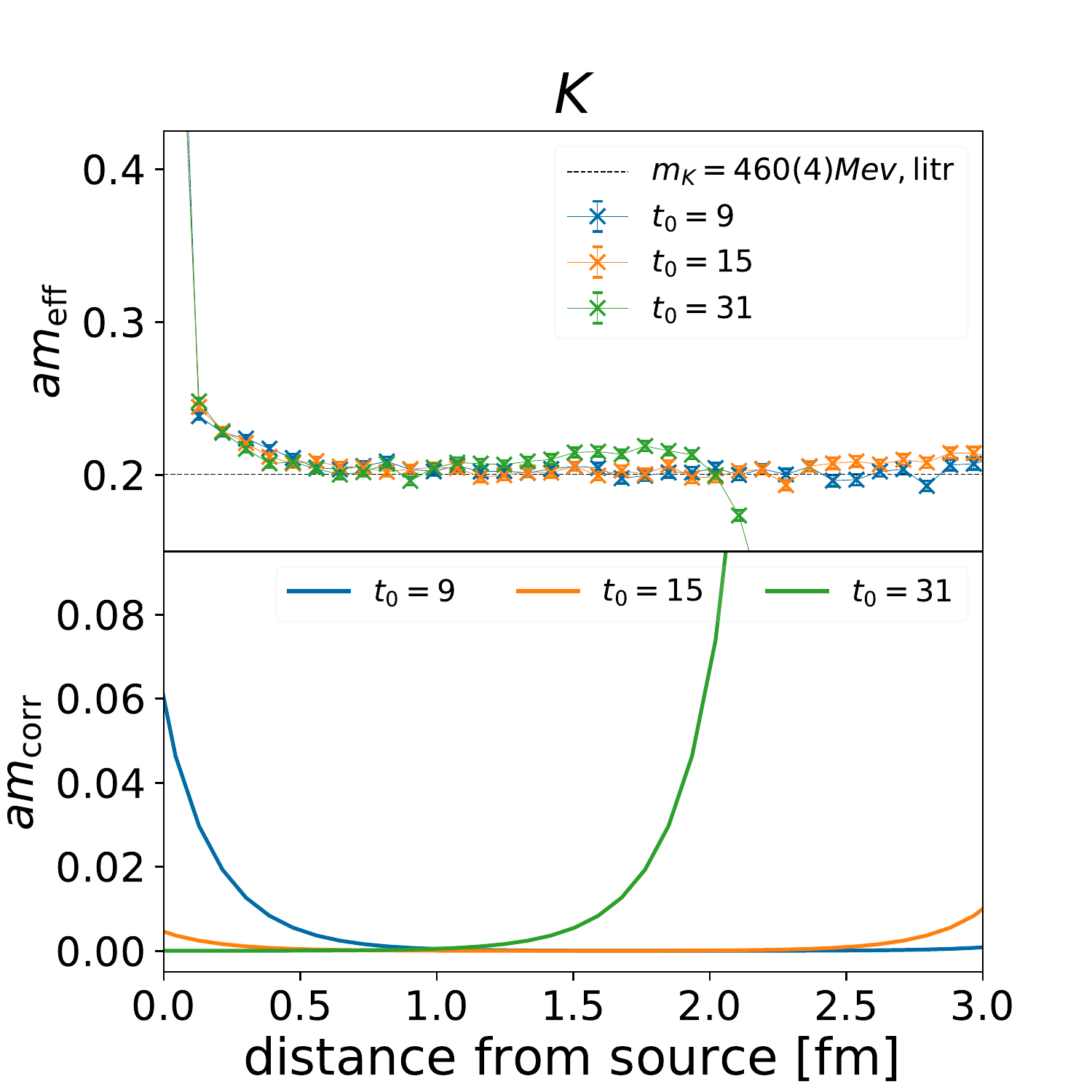}
    \includegraphics[width=0.32\textwidth]{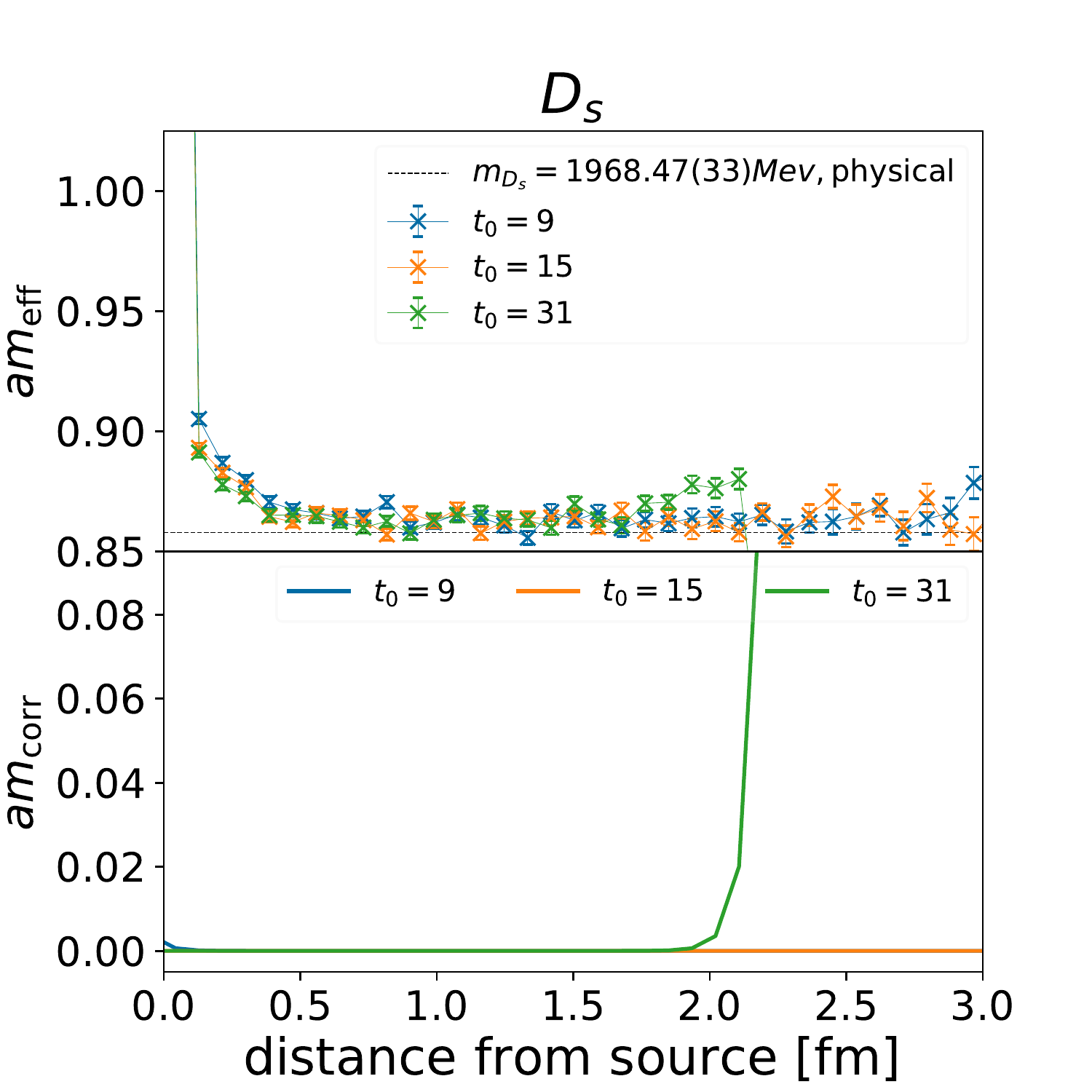}
    \caption{Comparison of the corrected effective masses (top) for $\pi$, $K$
      and $D_s$ respectively, and the corresponding cumulative correction
      $m_{\textrm{corr}}(t) + m_{\textrm{corr}}(T - t)$ (bottom).}
    \label{fig:meff_corr}
\end{figure}

This quantifies the amount of contamination stemming from the presence of the
open boundary conditions.  Since we are interested into the corrections coming
from both open boundaries we correct our numerical estimate for the effective
mass ($m_{\log}(t)$) with $m_{\textrm{corr}}$ and its time reverse
\begin{equation}
    m_{\textrm{eff}}(t) = m_{\log}(t) - m_{\textrm{corr}}(t) - m_{\textrm{corr}}(T - t).
\end{equation}
As shown in figure~\ref{fig:meff_corr}, this procedure is able to substantially
reduce the boundary effects on the effective masses. For each meson, we
extracted the two parameters $\tilde{T}$ and $m$ by simultaneously fitting the
two longer branches of the meson correlation functions generated from the
sources at $t_0 = 9$ and $15$. In all cases we obtain values of the reduced
$\chi^2$ lower that unity. While some deviations are still visible close to
$\tilde{T}$, the corrected effective masses show clear plateaux for a wide range
of time separations. This fitting procedure can be easily adapted to the
evaluation of the ratios of two-- and three--point correlation functions
required for the study of semileptonic decays.

%%%%%%%%%%%%%%%%%%%%%%%%%%%%%%%%%%%%%%%%%%%%%%%%%%%%%%%%
%%%%    conclusions                                 %%%%
%%%%%%%%%%%%%%%%%%%%%%%%%%%%%%%%%%%%%%%%%%%%%%%%%%%%%%%%
\section{Conclusions and future developments}
We have performed an extensive analysis of the effects of open boundary
conditions on the two--point correlation function computed with the LapH
formalism. On the ensemble we analysed and for our given statistical
uncertainties, we found that the region affected by the boundary extends into
the lattice for $1$ fm (12 time--slices), leaving us with more than half of the
lattice where the time translational invariance is restored. Our two--point
functions analysis lead to the identification of an optimal setup fixing the
source position at $t_0\sim15$ and number of eigenvectors to $N_{ev}\sim24$.

We further devised a fitting procedure that allows for the estimate of the
boundary contribution to the effective masses and a possible correction scheme
that has proven to greatly reduce the boundary effects on the meson effective
masses and that is largely independent of the source/sink positions.

Moreover, we have started generating data for three--point functions with local
current insertions. Using to the distillation workflow this will allow for the
evaluation of the processes needed to study CKM matrix elements and semileptonic
form factors with minimal computational overhead.

%%%%%%%%%%%%%%%%%%%%%%%%%%%%%%%%%%%%%%%%%%%%%%%%%%%%%%%%
%%%%    acknowledgments                             %%%%
%%%%%%%%%%%%%%%%%%%%%%%%%%%%%%%%%%%%%%%%%%%%%%%%%%%%%%%%
\section{Acknowledgments}

The work of J.T.T. has received funding from the European Union's Horizon 2020
research and innovation programme under the Marie Sk{\l}odowska-Curie grant
agreement No 894103. The DFF Research project 1. Grant n. 8021-00122B fully
supports the work of O.F. and partially supports M.D.M and
J.T.T. work. Computations were performed on UCloud interactive HPC system and
ABACUS2.0 supercomputer, which are managed by the eScience Center at the
University of Southern Denmark.

%%%%%%%%%%%%%%%%%%%%%%%%%%%%%%%%%%%%%%%%%%%%%%%%%%%%%%%%
%%%%    bibliography                                %%%%
%%%%%%%%%%%%%%%%%%%%%%%%%%%%%%%%%%%%%%%%%%%%%%%%%%%%%%%%
\bibliographystyle{JHEP}
\bibliography{lattice2022}

\providecommand{\href}[2]{#2}\begingroup\raggedright\begin{thebibliography}{10}

\bibitem{HadronSpectrum:2009krc}
{\scshape Hadron Spectrum} collaboration, \emph{{A Novel quark-field creation
  operator construction for hadronic physics in lattice QCD}},
  \href{https://doi.org/10.1103/PhysRevD.80.054506}{\emph{Phys. Rev. D}
  {\bfseries 80} (2009) 054506}
  [\href{https://arxiv.org/abs/0905.2160}{{\ttfamily 0905.2160}}].

\bibitem{Boyle:2019cdl}
P.~Boyle, F.~Erben, M.~Marshall, F.~O. H\'og\'ain, A.~Portelli and J.~T. Tsang,
  \emph{{An exploratory study of heavy-light semileptonic form factors using
  distillation}}, \href{https://doi.org/10.22323/1.363.0169}{\emph{PoS}
  {\bfseries LATTICE2019} (2019) 169}
  [\href{https://arxiv.org/abs/1912.07563}{{\ttfamily 1912.07563}}].

\bibitem{Cabibbo:1963yz}
N.~Cabibbo, \emph{Unitary symmetry and leptonic decays},
  \href{https://doi.org/10.1103/PhysRevLett.10.531}{\emph{Phys. Rev. Lett.}
  {\bfseries 10} (1963) 531}.

\bibitem{Kobayashi:1973fv}
M.~Kobayashi and T.~Maskawa, \emph{{CP Violation in the Renormalizable Theory
  of Weak Interaction}}, \href{https://doi.org/10.1143/PTP.49.652}{\emph{Prog.
  Theor. Phys.} {\bfseries 49} (1973) 652}.

\bibitem{London:2021lfn}
D.~London and J.~Matias, \emph{{$B$ Flavour Anomalies: 2021 Theoretical Status
  Report}},
  \href{https://doi.org/10.1146/annurev-nucl-102020-090209}{\emph{Ann. Rev.
  Nucl. Part. Sci.} {\bfseries 72} (2022) 37}
  [\href{https://arxiv.org/abs/2110.13270}{{\ttfamily 2110.13270}}].

\bibitem{Morningstar:2011ka}
C.~Morningstar, J.~Bulava, J.~Foley, K.~J. Juge, D.~Lenkner, M.~Peardon et~al.,
  \emph{{Improved stochastic estimation of quark propagation with Laplacian
  Heaviside smearing in lattice QCD}},
  \href{https://doi.org/10.1103/PhysRevD.83.114505}{\emph{Phys. Rev. D}
  {\bfseries 83} (2011) 114505}
  [\href{https://arxiv.org/abs/1104.3870}{{\ttfamily 1104.3870}}].

\bibitem{Luscher:2012av}
M.~Luscher and S.~Schaefer, \emph{{Lattice QCD with open boundary conditions
  and twisted-mass reweighting}},
  \href{https://doi.org/10.1016/j.cpc.2012.10.003}{\emph{Comput. Phys. Commun.}
  {\bfseries 184} (2013) 519}
  [\href{https://arxiv.org/abs/1206.2809}{{\ttfamily 1206.2809}}].

\bibitem{Bruno:2014jqa}
M.~Bruno et~al., \emph{{Simulation of QCD with N$_{f} =$ 2 $+$ 1 flavors of
  non-perturbatively improved Wilson fermions}},
  \href{https://doi.org/10.1007/JHEP02(2015)043}{\emph{JHEP} {\bfseries 02}
  (2015) 043} [\href{https://arxiv.org/abs/1411.3982}{{\ttfamily 1411.3982}}].

\bibitem{Gerardin:2019rua}
A.~G\'erardin, M.~C\`e, G.~von Hippel, B.~H\"orz, H.~B. Meyer, D.~Mohler
  et~al., \emph{{The leading hadronic contribution to $(g-2)_\mu$ from lattice
  QCD with $N_{\rm f}=2+1$ flavours of O($a$) improved Wilson quarks}},
  \href{https://doi.org/10.1103/PhysRevD.100.014510}{\emph{Phys. Rev. D}
  {\bfseries 100} (2019) 014510}
  [\href{https://arxiv.org/abs/1904.03120}{{\ttfamily 1904.03120}}].

\bibitem{Guagnelli:1999zf}
{\scshape ALPHA} collaboration, \emph{{Hadron masses and matrix elements from
  the QCD Schrodinger functional}},
  \href{https://doi.org/10.1016/S0550-3213(99)00466-6}{\emph{Nucl. Phys. B}
  {\bfseries 560} (1999) 465}
  [\href{https://arxiv.org/abs/hep-lat/9903040}{{\ttfamily hep-lat/9903040}}].

\end{thebibliography}\endgroup

\end{document}